\documentclass[11pt]{article}
\usepackage{jheppub}

\newcommand\fft[2]{{\frac{#1}{#2}}}
\newcommand\ft[2]{{\textstyle\frac{#1}{#2}}}
\newcommand\nn{\nonumber}

\preprint{MCTP-13-08}

\title{\boldmath The spectrum of IIB supergravity on $\mathrm{AdS}_5\times S^5/\mathbb Z_3$
and a $1/N^2$ test of AdS/CFT}

\author[a]{Arash Arabi Ardehali,}
\author[a]{James T. Liu,}
\author[b]{Phillip Szepietowski}

\affiliation[a]{Michigan Center for Theoretical Physics, Randall Laboratory of Physics,\\
The University of Michigan, Ann Arbor, MI 48109--1040, USA}
\affiliation[b]{Department of Physics, University of Virginia,\\
Box 400714, Charlottesville, VA 22904, USA}

\emailAdd{ardehali@umich.edu}
\emailAdd{jimliu@umich.edu}
\emailAdd{pgs8b@virginia.edu}

\abstract{We present the complete Kaluza-Klein spectrum resulting from the
compactification of IIB supergravity on $S^5/\mathbb Z_3$.  Knowledge of this
spectrum allows us to perform a holographic computation of the difference of
central charges $c-a$ of the dual $\mathrm{SU}(N)^3$ quiver gauge theory.
We find the numerical value $c-a=3/16$, in exact agreement with the field
theory result.}

\begin{document}
\maketitle
\flushbottom

\section{Introduction}
\label{sec:intro}

As a strong/weak coupling duality, AdS/CFT allows us to gain additional insight into the
non-perturbative regime of gauge theories, and in particular supersymmetric Yang-Mills
theories.  For the same reason, however, it is often a challenge to compare quantities on
both sides of the duality.  Even when a calculation can be done on both sides, the
results do not necessarily agree.  A well known example of this is the result for
the free energy of $\mathcal N=4$ SYM in going between weak coupling and strong coupling.
While both sides of the duality agree on the general behavior $F\sim N^2T^4$ (as expected
from a large $N$ conformal theory), the free energy picks up a $3/4$ factor in going from
weak to strong coupling \cite{Gubser:1996de,Gubser:1998nz}.

Of course, there is no reason to expect the free energy to remain independent of the
coupling.  However, direct comparisons can be made between protected quantities,
and there has been much exploration in this direction.  In this context, the Weyl anomaly
has proven useful in making the connection between field theories and their holographic
duals.  In four dimensions, the Weyl anomaly manifests itself in the trace of the stress tensor
\begin{equation}
\langle T^\mu_\mu\rangle=\fft1{16\pi^2}(cC_{\mu\nu\rho\sigma}^2-aE_4),
\label{eq:canda}
\end{equation}
where $C_{\mu\nu\rho\sigma}$ is the Weyl tensor, and $E_4=R_{\mu\nu\rho\sigma}^2
-4R_{\mu\nu}^2+R$ is the four-dimensional Euler density.  Here there are two central
charges, $a$ and $c$, and the former has received much recent attention as the subject
of the $a$-theorem in four \cite{Cardy:1988cwa,Komargodski:2011vj} and possibly higher
\cite{Elvang:2012st,Elvang:2012yc} dimensions.

Consider the case of $\mathcal N=4$ SYM with gauge group $\mathrm{SU}(N)$.
A one-loop computation in the field theory gives
\begin{equation}
c=a=\fft{N^2-1}4,
\end{equation}
where $N^2-1$ is the dimension of $\mathrm{SU}(N)$ and represents the weakly
coupled degrees of freedom.  The dual theory is given by IIB string theory on
$\mathrm{AdS}_5\times S^5$, and the holographic Weyl anomaly, first computed in
\cite{Henningson:1998gx}, matches the field theory result at leading order in $N^2$.
More generally, for IIB string theory on $\mathrm{AdS}_5\times X^5$, the leading
contribution is obtained from the classical gravity sector, and gives \cite{Gubser:1998vd}
\begin{equation}
c=a=\fft{N^2}4\fft{\pi^3}{\mathrm{vol}(X^5)}.
\label{eq:holoweyl}
\end{equation}

Our present interest is to extend this comparison between the field theory and the
holographic calculation beyond leading order in $N^2$.  For the gravitational dual to
$\mathcal N=4$ SYM, the $\mathcal O(1)$ subtraction that shifts $N^2\to N^2-1$ was obtained
in \cite{Bilal:1999ph,Bilal:1999ty,Mansfield:2000zw,Mansfield:2002pa,Mansfield:2003gs}
by summing over one-loop corrections due to the Kaluza-Klein tower on $S^5$.  Moreover,
in theories with reduced supersymmetry there are additional potential sources
for subleading contributions to $c$ and $a$.  Since $c\sim a\sim N^2$ at leading order,
it is often useful to characterize these corrections as a shift in $c-a$.  Open string loops,
and in particular the inclusion of $D$-branes (such as $D7$ flavor branes) will induce a
$1/N$ correction, while closed string loops will induce a $1/N^2$ correction.

In this paper, we focus on IIB string theory on the $S^5/\mathbb Z_3$ orbifold.  This orbifolding
reduces the supersymmetry, and as a result the dual quiver gauge theory is $\mathcal N=1$
SYM with gauge group $\mathrm{SU}(N)^3$.  On the gauge theory side, it is easy to see that the
$\mathcal O(1)$ contribution to $c-a$ is simply
\begin{equation}
c-a=\fft3{16},
\label{eq:316}
\end{equation}
corresponding to three decoupled $\mathcal N=1$ vector multiplets from the three $\mathrm U(1)$
factors.  Our aim is to reproduce this $3/16$ factor from the gravity side of the duality.  Since
$\mathbb Z_3$ acts freely on $S^5$, the $S^5/\mathbb Z_3$ orbifold has no singularities, and
hence requires no brane sources.  As a result, the first correction to $c-a$ arises from closed
string loops, in agreement with $c-a$ being an $\mathcal O(1)$ effect.

There are, in fact, two sources of closed string loop corrections.  The first originates from massive
string states in the loop, and is most directly encoded by the one-loop $R^4$ term in the type
II string effective action in ten dimensions
\cite{Sakai:1986bi,Vafa:1995fj,Duff:1995wd,Anselmi:1998zb}.  The
compactification of the $R^4$ correction on a Sasaki-Einstein manifold $\mathrm{SE}_5$
was investigated in \cite{Liu:2010gz}, and the result is that the five-dimensional action will
pick up a $R^2$ term of the form
\begin{equation}
S = \frac{1}{2\kappa_5^2}\int d^5x \left(R + \fft{12}{L^2}
+ \alpha R_{\mu\nu\rho\sigma}R^{\mu\nu\rho\sigma}+\cdots\right),
\end{equation}
where the coefficient $\alpha$ may be determined in terms of the data specifying
$\mathrm{SE}_5$ when written in canonical form as $\mathrm U(1)$ fibered over a
Kahler-Einstein base.  The inclusion of this Riemann-squared correction modifies
the holographic Weyl anomaly computation \cite{Nojiri:1999mh,Blau:1999vz}, and
we find
\begin{equation}
c-a=\fft\alpha{8L^2}a_0,\quad\mbox{where}\quad
a_0=\fft{\pi^2L^3}{\kappa_5^2}.
\label{eq:acc}
\end{equation}
(Here $a_0$ denotes the leading order central charge.)
Note, however, that this contribution from massive string loops vanishes
in the case of $S^5/\mathbb Z_3$, as the geometry is locally that of
$S^5$, which has $\alpha=0$.

Since massive string loops do not contribute to $c-a$ for the $S^5/\mathbb Z_3$
orbifold, the factor (\ref{eq:316}) must come entirely from the second type of correction.
This correction arises at the one-loop level with particles in the Kaluza-Klein tower running
in the loop.  As mentioned above, this loop correction was previously computed in order to
demonstrate the shift $N^2\to N^2-1$ in both $a$ and $c$ for IIB supergravity on $S^5$.
Thus all that is
needed here is to repeat the procedure, but this time with the spectrum of IIB supergravity
on $S^5/\mathbb Z_3$.  We follow the approach of
\cite{Mansfield:2000zw,Mansfield:2002pa,Mansfield:2003gs}, and in particular we
use the expression for the correction to the leading order Weyl anomaly
\begin{equation}
\delta\langle T^\mu_\mu\rangle=-\sum\fft{(E_0-2)a_2}{32\pi^2},
\end{equation}
where the sum is over all states in the KK tower.  Here $E_0$ is the lowest energy
defining the representation and $a_2$ is a four-dimensional heat kernel coefficient
(with an extra sign for anti-commuting fields).  Comparing this with (\ref{eq:canda}) gives
\begin{equation}
c-a=-\fft12\sum(E_0-2)a_2\Big|_{R_{\mu\nu\rho\sigma}^2\;\rm term},
\label{eq:MNU}
\end{equation}
where, since  $c=a$ at leading order for $S^5/\mathbb Z_3$, the entire
contribution to $c-a$ is from the Kaluza-Klein spectrum.

Our main result is that the sum over Kaluza-Klein modes in (\ref{eq:MNU}) gives
$3/16$, and hence exactly matches the field theory result.  In order to perform the
sum, we of course need the KK spectrum on $S^5/\mathbb Z_3$, which may be
obtained by $\mathbb Z_3$ projection of the $S^5$ spectrum.  As a bonus,
we also elucidate the $\mathcal N=2$ multiplet structure and shortening conditions
of this spectrum.

In section~\ref{sec:KK}, we examine the KK spectrum of IIB supergravity compactified
on $S^5/\mathbb Z_3$.  Then, in section~\ref{sec:Weyl}, we compute $c-a$ by summing
over this spectrum and find perfect agreement with the dual quiver gauge theory.  Some
of the details of constructing the KK spectrum are relegated to appendices.

\section{The Kaluza-Klein spectrum of IIB supergravity on $\mathrm{AdS}_5\times S^5/\mathbb Z_3$}
\label{sec:KK}

Type IIB supergravity compactified on $S^5$ gives rise to $\mathcal N=8$ gauged supergravity
in five dimensions along with an infinite tower of Kaluza-Klein modes.  The KK spectrum was
worked out in \cite{Gunaydin:1984fk,Kim:1985ez} by expanding the linearized ten-dimensional
fields in harmonics on the five-sphere.  The harmonics fall into representations of the
$\mathrm{SU}(4)$ isometry group of $S^5$, and the KK levels may be labeled by a single integer
$p\ge2,$ which corresponds to the oscillator number in \cite{Gunaydin:1984fk}.  At each level
$p$, the fluctuations assemble into unitary representations of the supergroup $\mathrm{SU}(2,2|4)$.
Level $p=2$ corresponds to the massless $\mathcal N=8$ supergravity multiplet, level $p=3$ is shortened,
and levels $p\ge4$ are long $\mathcal N=8$ supermultiplets.

The bosonic subgroup of $\mathrm{SU}(2,2|4)$ is $\mathrm{SO}(2,4)\times\mathrm{SU}(4)$,
where $\mathrm{SO}(2,4)$ is the isometry group of $\mathrm{AdS}_5$ and $\mathrm{SU}(4)$
is the isometry group of $S^5$.  We label $\mathrm{AdS}_5$ representations by $D(E_0,s_1,s_2)$,
with $E_0$, $s_1$ and $s_2$ the quantum numbers of the lowest state under the maximal
compact subgroup $\mathrm{SO}(2)\times \mathrm{SU}(2)\times \mathrm{SU}(2)
\simeq \mathrm{SO}(2)\times \mathrm{SO}(4)\subset \mathrm{SO}(2,4)$.  For the KK spectrum on
$\mathrm{AdS}_5\times S^5$, since each state in $\mathrm{AdS}_5$ transforms under a
representation of the $R$-symmetry group $\mathrm{SU}(4)$, we also append the Dynkin labels
for the $\mathrm{SU}(4)$ representation, and label $\mathrm{AdS}_5\times S^5$ representations
as $D(E_0,s_1,s_2;l_1,l_2,l_3)$.  The KK spectrum is summarized in Table~\ref{tbl:S5spectrum}.

\begin{table}[tp]
\centering
\begin{tabular}{|l|l|l|}
\hline
Field&Representation&KK level\\
\hline
$\varphi^{(1)}$&$D(p,0,0;0,p,0)$&$p\ge2$\\
$\lambda^{(1)}$&$D(p+\ft12,\ft12,0;0,p-1,1)+D(p+\ft12,0,\ft12;1,p-1,0)$&$p\ge2$\\
$A_{\mu\nu}^{(1)}$&$D(p+1,1,0;0,p-1,0)+D(p+1,0,1;0,p-1,0)$&$p\ge2$\\
\hline
$\varphi^{(2)}$&$D(p+1,0,0;0,p-2,2)+D(p+1,0,0;2,p-2,0)$&$p\ge2$\\
$\varphi^{(3)}$&$D(p+2,0,0;0,p-2,0)+D(p+2,0,0;0,p-2,0)$&$p\ge2$\\
$\lambda^{(2)}$&$D(p+\ft32,\ft12,0;0,p-2,1)+D(p+\ft32,0,\ft12;1,p-2,0)$&$p\ge2$\\
$A_\mu^{(1)}$&$D(p+1,\ft12,\ft12;1,p-2,1)$&$p\ge2$\\
$\psi_\mu^{(1)}$&$D(p+\ft32,1,\ft12;1,p-2,0)+D(p+\ft32,\ft12,1;0,p-2,1)$&$p\ge2$\\
$h_{\mu\nu}$&$D(p+2,1,1;0,p-2,0)$&$p\ge2$\\
\hline
$\lambda^{(3)}$&$D(p+\ft32,\ft12,0;2,p-3,1)+D(p+\ft32,0,\ft12;1,p-3,2)$&$p\ge3$\\
$\lambda^{(4)}$&$D(p+\ft52,\ft12,0;0,p-3,1)+D(p+\ft52,0,\ft12;1,p-3,0)$&$p\ge3$\\
$A_\mu^{(2)}$&$D(p+2,\ft12,\ft12;1,p-3,1)+D(p+2,\ft12,\ft12;1,p-3,1)$&$p\ge3$\\
$A_{\mu\nu}^{(2)}$&$D(p+2,1,0;2,p-3,0)+D(p+2,0,1;0,p-3,2)$&$p\ge3$\\
$A_{\mu\nu}^{(3)}$&$D(p+3,1,0;0,p-3,0)+D(p+3,0,1;0,p-3,0)$&$p\ge3$\\
$\psi_\mu^{(2)}$&$D(p+\ft52,1,\ft12;1,p-3,0)+D(p+\ft52,\ft12,1;0,p-3,1)$&$p\ge3$\\
\hline
$\varphi^{(4)}$&$D(p+2,0,0;2,p-4,2)$&$p\ge4$\\
$\varphi^{(5)}$&$D(p+3,0,0;0,p-4,2)+D(p+3,0,0;2,p-4,0)$&$p\ge4$\\
$\varphi^{(6)}$&$D(p+4,0,0;0,p-4,0)$&$p\ge4$\\
$\lambda^{(5)}$&$D(p+\ft52,\ft12,0;2,p-4,1)+D(p+\ft52,0,\ft12;1,p-4,2)$&$p\ge4$\\
$\lambda^{(6)}$&$D(p+\ft72,\ft12,0;0,p-4,1)+D(p+\ft72,0,\ft12;1,p-4,0)$&$p\ge4$\\
$A_\mu^{(3)}$&$D(p+3,\ft12,\ft12;1,p-4,1)$&$p\ge4$\\
\hline
\end{tabular}
\caption{\label{tbl:S5spectrum} The Kaluza-Klein spectrum of IIB supergravity on
$\mathrm{AdS}_5\times S^5$.  The representations are labeled $D(E_0,s_1,s_2;l_1,l_2,l_3)$
where $(E_0,s_1,s_2)$ specifies the $\mathrm{SO(2,4)}$ $\mathrm{AdS}_5$ representation and
$(l_1,l_2,l_3)$ are the Dynkin labels of the $\mathrm{SU}(4)$ representation for $S^5$.  The
labeling of the fields correspond to that of Ref.~\cite{Gunaydin:1984fk}}
\end{table}

\subsection{The $S^5/\mathbb Z_3$ orbifold}

The $S^5/\mathbb Z_3$ orbifold is defined by the $\mathbb Z_3$ action
\begin{equation}
X^i\to e^{2\pi i/3} X^i,\qquad i=1,2,3,
\label{eq:Z3act}
\end{equation}
where the $X^i$ are complex coordinates on the transverse $\mathbb C^3$ space to the stack of
D3-branes.  Since this action is in the center of $\mathrm{SU}(3)$, the orbifold preserves
$\mathcal N=2$ supersymmetry in five dimensions and gives rise to an $\mathcal N=1$ quiver
gauge theory that was investigated in \cite{Kachru:1998ys,Lawrence:1998ja,Gukov:1998kn}.
Note that the action (\ref{eq:Z3act}) acts freely away from the origin, so that $S^5/\mathbb Z_3$ is
a lens space.  As a result, the KK spectrum on $S^5/\mathbb Z_3$ is given simply by the subset of
states on $S^5$ that are invariant under the $\mathbb Z_3$ action.  Determining the spectrum
thus reduces to an exercise in group theory that was initiated in \cite{Duff:1998us,Oz:1998hr}.
Here we complete this procedure and highlight the resulting $\mathcal N=2$ supermultiplet
structure.

To identify the KK states on $S^5/\mathbb Z_3$, we decompose $\mathrm{SU}(4)\supset
\mathrm{SU}(3)\times\mathrm U(1)$, where $\mathrm{SU}(4)$ is the isometry group of $S^5$,
and the remaining $\mathrm U(1)$ is the $\mathcal N=2$ $R$-symmetry.  We furthermore
normalize the $\mathrm U(1)$ charge by taking
\begin{equation}
\mathbf4 \to \mathbf3_{1/3} \oplus \mathbf1_{-1}.
\label{eq:U1norm}
\end{equation}
We now observe that the states invariant under (\ref{eq:Z3act}) are
those with triality zero. Thus what needs to be done is to take the
$\mathrm{SU}(4)$ representations in Table~\ref{tbl:S5spectrum},
branch them into $\mathrm{SU}(3)\times \mathrm U(1)$, and then
select the triality zero subset.  With the $\mathrm U(1)$
normalization given in (\ref{eq:U1norm}), it is easy to see that
this is equivalent to keeping only states with integer $R$-charge.

Obtaining the KK spectrum on $S^5/\mathbb Z_3$ is now a straightforward exercise in group theory,
and the result is presented in Table~\ref{tbl:S5Z3spectrum}.  The necessary branching rules and
details of the notation are given in Appendix~\ref{app:Z3trunc}.  Since each $\mathrm{SU}(4)$
representation branches into an entire series of $\mathrm{SU}(3)$ representations, we use the
compact notation $D(E_0,s_1,s_2;[a,b]_s)$ where the symbols $[a,b]_s$ are given by
\begin{equation}\label{eq:abintro}
[a,b]_s=\bigoplus_l(a+3l-p,b+2p-3l)_{2p-4l+s}.
\end{equation}
(We keep the KK level $p$ implicit for compactness of the notation.)
Additional subscripts inside the square brackets will be used to indicate
restrictions on the Dynkin labels, and are related to multiplet
shortening, as we will see below. For more details, see
Appendix~\ref{app:Z3trunc}.

\begin{table}[tp]
\centering
\begin{tabular}{|l|l|}
\hline
Field&Representation\\
\hline
$\varphi^{(1)}$&$D(p,0,0;[0,0]_0)$\\
$\lambda^{(1)}$&$D(p+\ft12,\ft12,0;[0,0_{>0}]_{-1}\oplus[1,-2]_{-1})
+D(p+\ft12,0,\ft12;[0_{>0},0]_1\oplus[-2,1]_1)$\\
$A_{\mu\nu}^{(1)}$&$D(p+1,1,0;[1,-2]_{-2})+D(p+1,0,1;[-2,1]_2)$\\
\hline
$\varphi^{(2)}$&$D(p+1,0,0;[0_{>1},0]_2\oplus[-2_{>0},1]_2\oplus[-4,2]_2)$\\
&$D(p+1,0,0;[0,0_{>1}]_{-2}\oplus[1,-2_{>0}]_{-2}\oplus[2,-4]_{-2})$\\
$\varphi^{(3)}$&$D(p+2,0,0;[-1,-1]_0)+D(p+2,0,0;[-1,-1]_0)$\\
$\lambda^{(2)}$&$D(p+\ft32,\ft12,0;[-2,1_{>0}]_1\oplus[-1,-1]_1)
+D(p+\ft32,0,\ft12;[1_{>0},-2]_{-1}\oplus[-1,-1]_{-1})$\\
$A_\mu^{(1)}$&$D(p+1,\ft12,\ft12;[1_{>0},-2]_0\oplus[0_{>0},0_{>0}]_0\oplus[-1,-1]_0
\oplus[-2,1_{>0}]_0)$\\
$\psi_\mu^{(1)}$&$D(p+\ft32,1,\ft12;[1_{>0},-2]_{-1}\oplus[-1,-1]_{-1})
+D(p+\ft32,\ft12,1;[-2,1_{>0}]_1\oplus[-1,-1]_1)$\\
$h_{\mu\nu}$&$D(p+2,1,1;[-1,-1]_0)$\\
\hline
$\lambda^{(3)}$&$D(p+\ft32,\ft12,0;[1_{>1},-2]_1\oplus[0_{>1},0_{>0}]_1\oplus[-1_{>0},-1]_1
\oplus[-2_{>0},1_{>0}]_0\oplus[-3,0]_1$\\
&\quad$\oplus[-4,2_{>0}]_1)
+D(p+\ft32,0,\ft12;[-2,1_{>1}]_{-1}\oplus[0_{>0},0_{>1}]_{-1}\oplus[-1,-1_{>0}]_{-1}$\\
&\quad$\oplus[1_{>0},-2_{>0}]_{-1}\oplus[0,-3]_{-1}\oplus[2_{>0},-4]_{-1})$\\
$\lambda^{(4)}$&$D(p+\ft52,\ft12,0;[-1,-1_{>0}]_{-1}\oplus[0,-3]_{-1})
+D(p+\ft52,0,\ft12;[-1_{>0},-1]_1\oplus[-3,0]_1)$\\
$A_\mu^{(2)}$&$D(p+2,\ft12,\ft12;[-1_{>0},-1]_2\oplus[-2_{>0},1_{>0}]_2\oplus[-3,0]_2
\oplus[-4,2_{>0}]_2)$\\
&$+D(p+2,\ft12,\ft12;[-1,-1_{>0}]_{-2}\oplus[1_{>0},-2_{>0}]_{-2}\oplus[0,-3]_{-2}
\oplus[2_{>0},-4]_{-2})$\\
$A_{\mu\nu}^{(2)}$&$D(p+2,1,0;[1_{>1},-2]_0\oplus[-1_{>0},-1]_0\oplus[-3,0]_0)$\\
&$+D(p+2,0,1;[-2,1_{>1}]_0\oplus[-1,-1_{>0}]_0\oplus[0,-3]_0)$\\
$A_{\mu\nu}^{(3)}$&$D(p+3,1,0;[-3,0]_2)+D(p+3,0,1;[0,-3]_{-2})$\\
$\psi_\mu^{(2)}$&$D(p+\ft52,1,\ft12;[-1_{>0},-1]_1\oplus[-3,0]_1)
+D(p+\ft52,\ft12,1;[-1,-1_{>0}]_{-1}\oplus[0,-3]_{-1})$\\
\hline
$\varphi^{(4)}$&$D(p+2,0,0;[2_{>1},-4]_0\oplus[0_{>0},-3]_0\oplus[1_{>1},-2_{>0}]_0
\oplus[0_{>1},0_{>1}]_0\oplus[-1_{>0},-1_{>0}]_0$\\
&\quad$\oplus[-2,-2]_0\oplus[-2_{>0},1_{>1}]_0\oplus[-3,0_{>0}]_0\oplus[-4,2_{>1}]_0)$\\
$\varphi^{(5)}$&$D(p+3,0,0;[-4,2_{>1}]_2\oplus[-3,0_{>0}]_2\oplus[-2,-2]_2)$\\
&$+D(p+3,0,0;[2_{>1},-4]_{-2}\oplus[0_{>0},-3]_{-2}\oplus[-2,-2]_{-2})$\\
$\varphi^{(6)}$&$D(p+4,0,0;[-2,-2]_0)$\\
$\lambda^{(5)}$&$D(p+\ft52,\ft12,0;[2_{>1},-4]_{-1}\oplus[1_{>1},-2_{>0}]_{-1}
\oplus[0_{>0},-3]_{-1}\oplus[-1_{>0},-1_{>0}]_{-1}$\\
&\quad$\oplus[-2,-2]_{-1}\oplus[-3,0_{>0}]_{-1})
+D(p+\ft52,0,\ft12;[-4,2_{>1}]_1\oplus[-2_{>0},1_{>1}]_1$\\
&\quad$\oplus[-3,0_{>0}]_1\oplus[-1_{>0},-1_{>0}]_1\oplus[-2,-2]_1\oplus[0_{>0},-3]_1)$\\
$\lambda^{(6)}$&$D(p+\ft72,\ft12,0;[-3,0_{>0}]_1\oplus[-2,-2]_1)
+D(p+\ft72,0,\ft12;[0_{>0},-3]_{-1}\oplus[-2,-2]_{-1})$\\
$A_\mu^{(3)}$&$D(p+3,\ft12,\ft12;[0_{>0},-3]_0\oplus[-1_{>0},-1_{>0}]_0\oplus[-2,-2]_0
\oplus[-3,0_{>0}]_0)$\\
\hline
\end{tabular}
\caption{\label{tbl:S5Z3spectrum} The Kaluza-Klein spectrum of IIB supergravity on
$\mathrm{AdS}_5\times S^5/\mathbb Z_3$.  The notation is explained in the text and in
Appendix~\ref{app:Z3trunc}.}
\end{table}

\subsection{Filling out $\mathcal N=2$ supermultiplets}

Since the $S^5/\mathbb Z_3$ orbifold preserves $\mathcal N=2$ supersymmetry in five
dimensions, the KK spectrum of Table~\ref{tbl:S5Z3spectrum} ought to fall into representations
of the $\mathcal N=2$ superalgebra $\mathrm{SU}(2,2|1)$.  These representations were
constructed in \cite{Flato:1983te,Dobrev:1985qv} (see also \cite{Freedman:1999gp}), and
may be labeled $\mathcal D(E_0,s_1,s_2;r)$, where the quantum numbers correspond to those
of the lowest energy state.  The generic long representations are given by
\begin{eqnarray}
\mathcal D(E_0,s_1,s_2;r)&=&D(E_0,s_1,s_2;r)\nn\\
&&+D(E_0+\ft12,s_1\pm\ft12,s_2;r-1)
+D(E_0+\ft12,s_1,s_2\pm\ft12,r+1)\nn\\
&&+D(E_0+1,s_1,s_2;r\pm2)+D(E_0+1,s_1\pm\ft12,s_2\pm\ft12,r)\nn\\
&&+D(E_0+\ft32,s_1\pm\ft12,s_2,r+1)+D(E_0+\ft32,s_1,s_2\pm\ft12,r-1)\nn\\
&&+D(E_0+2,s_1,s_2,r),
\end{eqnarray}
where the plus/minus signs are uncorrelated.  For particular values of the quantum numbers,
the long multiplets are truncated to shorter ones.  There are three multiplet shortening conditions
as follows:
\begin{eqnarray}
\mbox{conserved}:&&\qquad E_0=2+s_1+s_2,\nn\\
\mbox{chiral (antichiral)}:&&\qquad E_0=\ft32r\quad (E_0=-\ft32r),\nn\\
\mbox{semi-long I (semi-long II)}:&&\qquad E_0=2+2s_1-\ft32r\quad (E_0=2+2s_2+\ft32r).
\end{eqnarray}
All three possibilities will show up in the $S^5/\mathbb Z_3$ spectrum.

The $\mathcal N=2$ multiplet structure of IIB supergravity compactified on $T^{1,1}$ was
highlighted in \cite{Ceresole:1999zs,Ceresole:1999ht}, and subsequently it was demonstrated
in \cite{Eager:2012hx} that the same pattern of multiplets arise in any Sasaki-Einstein
compactification.  In the language of \cite{Ceresole:1999zs,Ceresole:1999ht}, the KK spectrum
arranges itself into nine families of supermultiplets: Graviton, Gravitino I through IV and Vector I
through IV.  (Other special multiplets or Betti multiplets may arise depending on topology, but
they are not present in the $S^5/\mathbb Z_3$ spectrum.)

The key to assembling the KK states into $\mathcal N=2$ multiplets is the realization that
all states in a given multiplet must transform in the same $\mathrm{SU}(3)$ representation.
Cursory examination of Table~\ref{tbl:S5Z3spectrum} indicates there are nine sets of
$\mathrm{SU}(3)$ representations:
\begin{equation}
[0,0],\quad
[1,-2],\quad
[-2,1],\quad
[-4,2],\quad
[2,-4],\quad
[-1,-1],\quad
[-3,0],\quad
[0,-3],\quad
[-2,-2],
\end{equation}
where we have suppressed the $R$-charge subscript.  However, this is somewhat misleading,
as the different symbols are not unique.  In particular, the second identity of (\ref{eq:symbident})
allows us to shift the two labels by three, so that there are only five distinct sets
\begin{equation}
[0,0],\quad
[1,-2]\sim[-2,1],\quad
[-4,2]\sim[-1,-1]\sim[2,-4],\quad
[-3,0]\sim[0,-3],\quad
[-2,-2].
\label{eq:fivesets}
\end{equation}
By examining the field content transforming in each of these sets, we may arrange the
states into the nine general families of \cite{Ceresole:1999zs,Ceresole:1999ht,Eager:2012hx}.
We find that the $[0,0]$ states make up Vector Multiplet I, the $[1,-2]\sim[-2,1]$ states fill out
Gravitino Multiplets I and III, the $[-4,2]\sim[-1,-1]\sim[2,-4]$ states split up into the Graviton
Multiplet, and Vector Multiplets III and IV, the $[-3,0]\sim[0,-3]$ states fill out Gravitino Multiplets
II and IV and the $[-2,-2]$ states correspond to Vector Multiplet II.  This is summarized in
Table~\ref{tbl:N=2spectrum}; for additional information, see Appendix~\ref{app:multiplets}.

\begin{table}[tp]
\centering
\begin{tabular}{|l|l|l|}
\hline
Supermultiplet&Representation&KK level\\
\hline
Graviton&$\mathcal D(p+1,\ft12,\ft12;[-1,-1]_0)$&$p\ge2$\\
Gravitino I and III&$\mathcal D(p+\ft12,\ft12,0;[1,-2]_{-1})+\mathcal D(p+\ft12,0,\ft12;[-2,1]_1)$&
$p\ge2$\\
Gravitino II and IV&$\mathcal D(p+\ft32,\ft12,0;[-3,0]_1)+\mathcal D(p+\ft32,0,\ft12;[0,-3]_{-1})$&
$p\ge3$\\
Vector I&$\mathcal D(p,0,0;[0,0]_0)$&$p\ge2$\\
Vector II&$\mathcal D(p+2,0,0;[-2,-2]_0)$&$p\ge4$\\
Vector III and IV&$\mathcal D(p+1,0,0;[-1,-1]_{-2})+\mathcal D(p+1,0,0;[-1,-1]_2)$&$p\ge2$\\
\hline
\end{tabular}
\caption{\label{tbl:N=2spectrum} The $\mathcal N=2$ spectrum of IIB supergravity on
$S^5/\mathbb Z_3$. In this table we employ a somewhat unorthodox notation in which the $SU(3)$ representation shorthand $[a,b]_s$ is written explicitly inside of each $\mathcal N = 2$ representation as $\mathcal D(E_0,s_1,s_2;[a,b]_s).$ This is due to the fact that different terms in the sum (\ref{eq:abintro}) have different $R$-charges. The notation should be interpreted as a sum of $\mathcal N = 2$ multiplets with each term having the appropriate $R$-charge and $SU(3)$ representation as the terms in the sum defined by $[a,b]_s.$}
\end{table}

\subsection{Multiplet shortening}

Until now, we have mostly ignored the subscripted constraints in the representation symbols
$[a_{>i},b_{>j}]_s$.  What these constraints indicate is that certain representations are absent
in the KK spectrum.  When arranged in supermultiplets, as in Table~\ref{tbl:N=2spectrum}, the
result is that some states are absent within supermultiplets.  It is natural to expect that this
corresponds to multiplet shortening, and in fact this is exactly what happens.  We list the
entire set of shortened multiplets in Table~\ref{tbl:KKshort}, and relegate the details to
Appendix~\ref{app:multiplets}.

\begin{table}[tp]
\centering
\begin{tabular}{|l|l|l|l|}
\hline
Multiplet&KK level&Shortened representation&Shortening type\\
\hline
Graviton&$p=2$&$\mathcal D(3,\ft12,\ft12;0)(0,0)$&conserved\\
&$p=3l+2$&$\mathcal D(3l+3,\ft12,\ft12;-2l)(3l,0)$&SLI\\
&&$\mathcal D(3l+3,\ft12,\ft12;2l)(0,3l)$&SLII\\
\hline
Gravitino I&$p=3l+1$&$\mathcal D(3l+\ft32,\ft12,0;2l+1)(0,3l)$&chiral \\
&& $\mathcal D(3l+\ft32,\ft12,0;-2l+1)(3l,0)$&SLI\\
&$p=3l+3$& $\mathcal D(3l+\ft72,\ft12,0;2l+1)(1,3l+1)$&SLII \\
\hline
Gravitino II&$p=3l+3$&$\mathcal D(3l+\ft92,\ft12,0;-2l-1)(3l,0)$&SLI\\
\hline
Gravitino III&$p=3l+1$ & $\mathcal D(3l+\ft32,0,\ft12;-2l-1)(3l,0)$&anti-chiral \\
&& $\mathcal D(3l+\ft32,0,\ft12;2l-1)(0,3l)$&SLII \\
&$p=3l+3$& $\mathcal D(3l+\ft72,0,\ft12;-2l-1)(3l+1,1)$ & SLI \\
\hline
Gravitino IV &$p=3l+3$& $\mathcal D(3l+\ft92,0,\ft12;2l+1)(0,3l)$&SLII\\
\hline
Vector I &$p=2$&$\mathcal D(2,0,0;0)(1,1)$&conserved\\
&$p=3l$&$\mathcal D(3l,0,0;2l)(0,3l)$&chiral\\
&&$\mathcal D(3l,0,0;-2l)(3l,0)$&anti-chiral\\
&$p=3l+2$&$\mathcal D(3l+2,0,0;-2l)(3l+1,1)$&SLI\\
&&$\mathcal D(3l+2,0,0;2l)(1,3l+1)$&SLII\\
\hline
Vector II & --- & --- & --- \\
\hline
Vector III &$p=3l+2$& $\mathcal D(3l+3,0,0;-2l-2)(3l,0)$& anti-chiral \\
&$p=3l+4$& $\mathcal D(3l+5,0,0;-2l-2)(3l+1,1)$& SLI \\
\hline
Vector IV &$p=3l+2$& $\mathcal D(3l+3,0,0;2l+2)(0,3l)$ & chiral \\
&$p=3l+4$& $\mathcal D(3l+5,0,0;2l+2)(1,3l+1)$ & SLII \\
\hline
\end{tabular}
\caption{\label{tbl:KKshort} Shortening structure of the $S^5/\mathbb Z_3$ KK tower.
The supermultiplets are given in the conventional notation $\mathcal D(E_0,s_1,s_2;r)$
with the $\mathrm{SU}(3)$ representation $(l_1,l_2)$ appended.  Note that Vector Multiplet
II is never shortened.}
\end{table}

Of particular interest are the multiplets arising at the non-generic $p=2$ and $p=3$ KK levels.
For $p=2$, the shortened multiplets are the Graviton (conserved), Vector I (conserved) and
Vectors III and IV (anti-chiral and chiral).  This corresponds to the gauged supergravity
sector, with the supergravity multiplet coupled to an $\mathrm{SU}(3)$ adjoint vector
multiplet and the universal hypermultiplet.  At the $p=3$ level, we encounter Gravitinos I and II
(semi-long) transforming in the adjoint, Gravitinos II and IV (semi-long), and a massive Vector I
(chiral and anti-chiral) transforming in the $\mathbf{10}$ and $\overline{\mathbf{10}}$ of
$\mathrm{SU}(3)$.  There are no long multiplets below $p=4$.

\subsection{Dual operators with protected dimension}

On the CFT side, the shortened multiplets are dual to superfields with protected dimension.
After having identified these multiplets, it is a fairly
straightforward task to find their dual superfields, building on the
earlier results of \cite{Oz:1998hr} and with the aid of the similar
expressions given for the Klebanov-Witten theory in \cite{Ceresole:1999zs}.
The quiver gauge theory consists
of $\mathrm{SU}(N)$ gauge superfields $V_{i}$, $i=1,2,3$, whose field strength
superfields we denote by $W^{\alpha}_{i}$, and three triplets of chiral
superfields $A_a, B_a, C_a$, $a=1,2,3$, transforming according to the $(N,\bar{N},1),\
(1,N,\bar{N}),\ (\bar{N},1,N)$ representations of the gauge group.
Now, the chiral multiplets in Vector Multiplet I, the chiral tensor
multiplets in Gravitino Multiplet I and the chiral multiplets in
Vector Multiplet IV correspond respectively to the chiral
superfields of the form%
\footnote{As in \cite{Ceresole:1999zs}, while we do not make it
explicit, we always mean the symmetrized trace (over the $a$
indices) and properly inserted field strengths. For example,
$T^{4}=Tr\left(W^\alpha_{1}A_{(a}B_{b}C_{c)}+A_{(a}W^\alpha_{2}B_{b}C_{c)}
+A_{(a}B_{b}W^\alpha_{3}C_{c)}\right)$.}
\begin{equation}
S^{p}=Tr\left((ABC)^{p/3}\right),\quad \Delta^{p}=p, \quad r=\frac{2}{3}p,\quad
p=3l\ge 3,
\end{equation}
\begin{equation}
T^{p}=Tr\left(W_{\alpha}(ABC)^{(p-1)/3}\right),\quad
\Delta^{p}=p+1/2, \quad r=\frac{2}{3}p+\frac{1}{3},\quad p=3l+1\ge4,
\end{equation}
\begin{equation}
\Phi^{p}=Tr\left(W^{\alpha}W_{\alpha}(ABC)^{(p-2)/3}\right),\quad
\Delta^{p}=p+1, \quad r=\frac{2}{3}p+\frac{2}{3},\quad p=3l+2\ge 2.
\end{equation}
The semi-long multiplets in Graviton Multiplet, Gravitino Multiplet
I (SLI), Gravitino Multiplet II (SLI), Gravitino Multiplet I (SLII),
Vector Multiplet I and Vector Multiplet IV correspond respectively
to the (semi-)conserved superfields of the form
\begin{equation}
J^{p}_{\alpha\dot{\alpha}}=Tr\left(J_{\alpha\dot{\alpha}}(ABC)^{(p-2)/3}\right),\quad
\Delta^{p}=p+1, \quad r=\frac{2}{3}p-\frac{4}{3},\quad p=3l+2\ge 2,
\end{equation}
\begin{equation}
L^{p}_{1\dot\alpha}=Tr\left(e^V \bar{W}_{\dot{\alpha}}e^{-V}(ABC)^{(p-1)/3}\right),\quad
\Delta^{p}=p+1/2, \quad r=-\frac{2}{3}p+\frac{5}{3},\quad p=3l+1\ge 4,
\end{equation}
\begin{equation}
L^{p}_{2\dot\alpha}=Tr\left(e^V \bar{W}_{\dot{\alpha}}e^{-V}W^2(ABC)^{(p-3)/3}\right),\quad
\Delta^{p}=p+3/2, \quad r=-\frac{2}{3}p+1,\quad p=3l+3\ge 3,
\end{equation}
\begin{equation}
L^{p}_{3\alpha}=Tr\left(W_{\alpha}(Ae^{V}\bar{A}e^{-V})(ABC)^{(p-3)/3}\right),\quad
\Delta^{p}=p+1/2, \quad r=\frac{2}{3}p-1,\quad p=3l + 3\ge 3,
\end{equation}
\begin{equation}
J^{p}=Tr\left(J(ABC)^{(p-2)/3}\right),\quad \Delta^{p}=p, \quad
r=\frac{2}{3}p-\frac{4}{3},\quad p=3l+2\ge 2,
\end{equation}
\begin{equation}
I^{p}=Tr\left(JW^{2}(ABC)^{(p-4)/3}\right),\quad \Delta^{p}=p+1,
\quad r=\frac{2}{3}p-\frac{2}{3},\quad p=3l+4\ge 4,
\end{equation}
where
\begin{equation}
J_{\alpha\dot{\alpha}}=W_{\alpha}e^{V}\bar{W}_{\dot{\alpha}}e^{-V},
\end{equation}
\begin{equation}
J=A(e^{V}\bar{A}e^{-V}).
\end{equation}
(Although we have singled out the chiral superfield $A$, the proper symmetrization
over the chiral superfields should be understood.)

Note in particular that CFT operators dual to AdS multiplets at KK
level $p$ have exactly $p$ superfields in them, similar to the case
of the $S^{5}$ compactification.  This is to be expected, of course, as
$S^5/\mathbb Z_3$ is simply related to $S^5$ by orbifolding.  This
connection between KK level and the length of the dual operators
will provide some insight into the regularization scheme used in the
following section.

\section{The holographic computation of $c-a$}
\label{sec:Weyl}

Before proceeding with the computation of $c-a$, it is worth reviewing the leading
order Weyl anomaly for the $S^5/\mathbb Z_3$ theory.  On the
gauge theory side, the $\mathrm{SU}(N)^3$ quiver contains three vector multiplets
($c=1/8$, $a=3/16$) in the adjoint and nine chiral multiplets ($c=1/24$, $a=1/48$) in
bifundamentals.  Summing up these contributions then gives
\begin{equation}
c=\fft{3N^2}4-\fft38,\qquad a=\fft{3N^2}4-\fft9{16},
\label{eq:SUN3}
\end{equation}
so that $c-a=3/16$.  On the gravity side, we use the leading order holographic
Weyl anomaly expression (\ref{eq:holoweyl}) with
$\mathrm{vol}(S^5/\mathbb Z_3)=\mathrm{vol}(S^5)/3$ to obtain
$c=a=3N^2/4$, which agrees with the above at $\mathcal O(N^2)$.

In order to obtain the $\mathcal O(1)$ contribution to $c-a$, we
need to sum over all states in the KK tower according to
(\ref{eq:MNU}).  Using the heat kernel coefficients in
\cite{Christensen:1978md}, we tabulate the contribution of
individual fields to $c-a$ in Table~\ref{tbl:hkc-a}.  Since we are
interested in representations of $\mathcal N=2$ supersymmetry, we
now sum these contributions over each component of the multiplet for
vector, gravitino and graviton multiplets.  The result is presented
in Table~\ref{tbl:shortc-a}.

Note in particular that long multiplets do not contribute to $c-a$,
so we only need to sum over the shortened spectrum for
$S^5/\mathbb Z_3$ as given in Table~\ref{tbl:KKshort}. In fact,
since equation (\ref{eq:MNU}) is derived in \cite{Mansfield:2003gs}
from relations involving bare masses of the bulk theory, the
vanishing contribution of long multiplets (with presumably
unprotected masses) is essential for the secured computability of
the subleading Weyl anomaly of the boundary theory. Overlooking this
subtlety one might have attempted to reproduce the individual
central charges $c$ and $a$ of the quiver gauge theory following
\cite{Mansfield:2002pa}, but then the contributions of long
multiplets become non-vanishing and a knowledge of renormalized
masses of the bulk theory is required.

\begin{table}[tp]
\centering
\begin{tabular}{|l|l|l|}
\hline
Field&Representation&Contribution to $360(c-a)$\\
\hline
$\phi$&$D(E_0,0,0)$&$-(E_0-2)$\\
$\lambda$&$D(E_0,\fft12,0)+D(E_0,0,\fft12)$&$-\fft72(E_0-2)$\\
$A_\mu$&$D(3,\fft12,\fft12)$&$13$\\
&$D(E_0>3,\fft12,\fft12)$&$11(E_0-2)$\\
$A_{\mu\nu}$&$D(E_0,1,0)+D(E_0,0,1)$&$-33(E_0-2)$\\
$\psi_\mu$&$D(\fft72,1,\fft12)+D(\fft72,\fft12,1)$&$173$\\
&$D(E_0>\fft72,1,\fft12)+D(E_0>\fft72,\fft12,1)$&$\fft{219}2(E_0-2)$\\
$h_{\mu\nu}$&$D(4,1,1)$&$-411$\\
&$D(E_0>4,1,1)$&$-189(E_0-2)$\\
\hline
\end{tabular}
\caption{\label{tbl:hkc-a} The contribution to $360(c-a)$ from fields with spins no higher than two.
The massless vector, gravitino and graviton contributions include the appropriate ghost sector.}
\end{table}

\begin{table}[tp]
\centering
\begin{tabular}{|l|c|c|c|}
\hline
&Vector&Gravitino&Graviton\\
&$\mathcal D(E_0,0,0;r)$&$\mathcal D(E_0,\fft12,0;r)$&$\mathcal D(E_0,\fft12,\fft12;r)$\\
\hline
Conserved&$\fft1{32}$&&$-\fft58$\\
Chiral&$-\fft1{96}(E_0-\fft32)$&$-\fft5{48}(E_0-\fft32)$&$\fft5{24}(E_0-\fft32)$\\
Anti-chiral&$-\fft1{96}(E_0-\fft32)$&&$\fft5{24}(E_0-\fft32)$\\
Semi-long I&$\fft1{96}(E_0-\fft12)$&$-\fft1{96}E_0$&$-\fft5{48}E_0$\\
Semi-long II&$\fft1{96}(E_0-\fft12)$&$\fft5{48}(E_0-\fft12)$&$-\fft5{48}E_0$\\
Long&$0$&$0$&$0$\\
\hline
\end{tabular}
\caption{\label{tbl:shortc-a} The contribution to $c-a$ from vector, gravitino and graviton multiplets.
Note the vanishing contribution from long multiplets.}
\end{table}

Writing out the sum over shortened multiplets for $S^5/\mathbb Z_3$, we find
\begin{equation}
c-a=\fft1{64}\sum_{p\ge2}\begin{cases}
p(-6p^2+3p+5),&p=2,5,8,\ldots\\
2p(6p^2-5),&p=3,6,9,\ldots\\
p(-6p^2-3p+5),&p=4,7,10,\ldots.
\end{cases}
\label{eq:c-asum}
\end{equation}
Note that this breaks up into three contributions based on $p\mod3$, as one may
expect from the nature of the $\mathbb Z_3$ orbifold.  As in the $S^5$ case treated in
\cite{Mansfield:2002pa}, this series is divergent, and hence needs to be regulated.
The regulation procedure used in \cite{Mansfield:2002pa} is to multiply each term in
the sum by $z^p$.  The sum then becomes absolutely convergent for $z<1$.  We then
analytically continue the result and examine the behavior as $z\to1$.  In the present
case, in fact, the result is finite for $z=1$, and we find $c-a=3/16$, thus matching the
gauge theory result.

Alternatively, we may perform a zeta function regularization.  This is complicated
somewhat by the fact that the sum splits into three expressions depending on
$p\mod3$.  Here it convenient to introduce the Hurwitz zeta function
\begin{equation}
\zeta(s,\alpha)=\sum_{n=0}^\infty(n+\alpha)^{-s},
\end{equation}
which generalizes the Riemann zeta function
\begin{equation}
\zeta(s)=\sum_{n=1}^\infty n^{-s}.
\end{equation}
We then break up the sum (\ref{eq:c-asum}) into three terms, with $p=3k+2$, $p=3k+3$
and $p=3k+4$
\begin{eqnarray}
c-a&=&\fft3{64}\sum_{k\ge0}\Bigl[
-54(k+\ft23)^3+9(k+\ft23)^2+5(k+\ft23)
+108(k+1)^3-10(k+1)\nn\\
&&\kern3.3em-54(k+\ft43)^3-9(k+\ft43)^2+5(k+\ft43)^2\Bigr]\nn\\
&=&\fft3{64}\Bigl[
-54\zeta(-3,\ft23)+9\zeta(-2,\ft23)+5\zeta(-1,\ft23)
+108\zeta(-3,1)-10\zeta(-1,1)\nn\\
&&\qquad-54\zeta(-3,\ft43)-9\zeta(-2,\ft43)+5\zeta(-1,\ft43)\Bigr]\nn\\
&=&\fft3{16},
\end{eqnarray}
in perfect agreement with the above.

\section{Discussion}
\label{sec:disc}

We have performed a one-loop test of AdS/CFT by matching the $\mathcal O(1)$
contribution to the difference of central charges, $c-a$, in both the $\mathrm{SU}(N)^3$
quiver gauge theory and its AdS dual.  In order to make this comparison, we have
explicitly obtained the KK spectrum of IIB supergravity on $S^5/\mathbb Z_3$.  As
expected on general grounds, the spectrum may be arranged into nine towers of
$\mathcal N=2$ supermultiplets in parallel with the $T^{1,1}$ case.

In addition to $S^5/\mathbb Z_3$, the $T^{1,1}$ case was considered in \cite{Liu:2010gz},
and a prediction was given that Kaluza-Klein loops ought to give a shift in $c-a$ of
$1/12$.  This would be in addition to the contribution $1/24$ that arises from massive
string loops.  Since the KK spectrum on $T^{1,1}$ is known
\cite{Ceresole:1999zs,Ceresole:1999ht}, it would be informative to see if this prediction
pans out.  This case is currently under investigation.

Finally, although we have focused on $c-a$, which is of $\mathcal O(1)$, it would be
desirable to reproduce either $c$ or $a$ directly, as given in (\ref{eq:SUN3}).  The
difficulty in doing so appears to be twofold.  Firstly, one would need to keep subleading
terms in the holographic computation of the $a$ central charge in (\ref{eq:acc}).  This
will involve higher derivative corrections to the volume of $S^5/\mathbb Z_3$ and the
effective AdS radius $L$.  Secondly, for the sum over KK states, while the heat kernel
coefficients leading to, say, $a$ are known, they will depend on Ricci terms that may
be shifted around in the one-loop determinants.  Thus additional care may be needed
to identify the appropriate equations of motion pertaining to the KK tower.

\acknowledgments

The idea to compute the subleading contribution to $c-a$ for IIB supergravity on
$\mathrm{AdS}_5\times S^5/\mathbb Z_3$ came out of discussions with R.~Minasian
following the completion of \cite{Liu:2010gz}. This work is supported in part by the US
Department of Energy under grants DE-SC0007859 and DE-SC0007984.

\vfill

\appendix

\section{The branching rules for $\mathrm{SU}_{4}\supset \mathrm{SU}_{3}\times
\mathrm U_{1}$ and projection onto $\mathbb Z_3$ singlets}
\label{app:Z3trunc}

The KK spectrum on $S^5/\mathbb Z_3$ is obtained by projecting the $S^5$ spectrum
onto $\mathbb Z_3$ invariant states.  This is done by first branching the relevant
$\mathrm{SU}(4)$ representations under
$\mathrm{SU}_{4}\supset \mathrm{SU}_{3}\times \mathrm U_{1}$ and then selecting
the triality zero representations of $\mathrm{SU}(3)$.

Based on the $\mathrm{SU}(4)$ representations in Table~\ref{tbl:S5spectrum}, we need the
following branching rules:
\begin{eqnarray}
(0,n,0)&=&\bigoplus_{k=0}^n(k,n-k)_{(2n-4k)/3},\nn\\
(1,n,0)&=&\bigoplus_{k=0}^n(k+1,n-k)_{(2n-4k+1)/3}\bigoplus_{k=0}^n(k,n-k)_{(2n-4k-3)/3},\nn\\
(2,n,0)&=&\bigoplus_{k=0}^n(k+2,n-k)_{(2n-4k+2)/3}\bigoplus_{k=0}^n(k+1,n-k)_{(2n-4k-2)/3}
\bigoplus_{k=0}^n(k,n-k)_{(2n-4k-6)/3},\nn\\
(1,n,1)&=&\bigoplus_{k=0}^n(k+1,n-k)_{(2n-4k+4)/3}\bigoplus_{k=0}^n(k+1,n-k+1)_{(2n-4k)/3}
\bigoplus_{k=0}^n(k,n-k)_{(2n-4k)/3}\nn\\
&&\bigoplus_{k=0}^n(k,n-k+1)_{(2n-4k-4)/3},\nn\\
(2,n,1)&=&\bigoplus_{k=0}^n(k+2,n-k)_{(2n-4k+5)/3}\bigoplus_{k=0}^n(k+2,n-k+1)_{(2n-4k+1)/3}
\nn\\
&&\bigoplus_{k=0}^n(k+1,n-k)_{(2n-4k+1)/3}\bigoplus_{k=0}^n(k+1,n-k+1)_{(2n-4k-3)/3}\nn\\
&&\bigoplus_{k=0}^n(k,n-k)_{(2n-4k-3)/3}\bigoplus_{k=0}^n(k,n-k+1)_{(2n-4k-7)/3},\nn\\
(2,n,2)&=&\bigoplus_{k=0}^n(k+2,n-k)_{(2n-4k+8)/3}\bigoplus_{k=0}^n(k+1,n-k)_{(2n-4k+4)/3}
\nn\\
&&\bigoplus_{k=0}^n(k+2,n-k+1)_{(2n-4k+4)/3}\bigoplus_{k=0}^n(k+2,n-k+2)_{(2n-4k)/3}\nn\\
&&\bigoplus_{k=0}^n(k+1,n-k+1)_{(2n-4k)/3}\bigoplus_{k=0}^n(k,n-k)_{(2n-4k)/3}\nn\\
&&\bigoplus_{k=0}^n(k+1,n-k+2)_{(2n-4k-4)/3}\bigoplus_{k=0}^n(k,n-k+1)_{(2n-4k-4)/3}\nn\\
&&\bigoplus_{k=0}^n(k,n-k+2)_{(2n-4k-8)/3}.
\label{eq:branching}
\end{eqnarray}
The representations are given by their Dynkin labels and the $\mathrm U(1)$ charge is
normalized by (\ref{eq:U1norm}).

Since a given $\mathrm{SU}(3)$ representation labeled by $(l_1,l_2)$ has triality $l_1+2l_2\mod3$,
the $\mathbb Z_3$ singlet states are those with $l_1+2l_2\equiv0\mod3$, or equivalently
$l_1\equiv l_2\mod3$.  Note that such states also have integer $R$-charge.  It is now a straightforward
exercise to obtain the $\mathbb Z_3$ singlets in the decomposition of the $\mathrm{SU}(4)$
representations given in (\ref{eq:branching}).  For example, the $(0,n,0)$ representation branches
into a sum of $\mathrm{SU}(3)$ representations with $R$-charge $(2n-4k)/3$.  The requirement
that this is an integer gives the condition $2n-4k=0\mod3$, which is equivalent to $n+k=0\mod3$.
We thus let $k=3l-n$ with $l\in\mathbb Z$, and find
\begin{equation}
(0,n,0)\longrightarrow\bigoplus_{l=\lceil n/3\rceil}^{\lfloor 2n/3\rfloor}(3l-n,2n-3l)_{2n-4l},
\label{eq:0n0singlet}
\end{equation}
under the $\mathbb Z_3$ projection.  The other representations in (\ref{eq:branching}) follow a
similar pattern.

We find it convenient to introduce a shorthand notation for sets of triality zero representations
that show up in the right hand side of expressions such as (\ref{eq:0n0singlet}).  Let
\begin{equation}
[a,b]_s(n)\equiv\bigoplus_{l=\lceil(n-a)/3\rceil}^{\lfloor(2n+b)/3\rfloor}(a+3l-n,b+2n-3l)_{2n-4l+s}.
\label{eq:abnotation}
\end{equation}
Note that the allowed values of $l$ in the sum are those for which the Dynkin labels give rise
to valid $\mathrm{SU}(3)$ representations.  In particular, the restriction is that $l_1\ge0$ and
$l_2\ge0$ for a representation labeled by $(l_1,l_2)$.  The symbols $[a,b]_s(n)$ satisfy the
following relations
\begin{eqnarray}
[a,b]_s(n-k)&=&[a+k,b-2k]_{s-2k}(n),\nn\\
{}[a,b]_s(n)&=&[a+3k,b-3k]_{s-4k}(n),\nn\\
\overline{[a,b]_s(n)}&=&[b,a]_{-s}(n),
\label{eq:symbident}
\end{eqnarray}
where the last line corresponds to the conjugate representation.  Triality zero states correspond
to $a\equiv b\mod3$.

Because of the small integer offsets such as $k+1$ or $n-k+1$ that show up in (\ref{eq:branching}),
in some cases it is necessary to impose a stronger restriction on allowed values of $l_1$ and $l_2$.
We thus allow for a refinement of the notation in (\ref{eq:abnotation}) by introducing
\begin{equation}
[a_{>i},b_{>j}]_s(n).
\end{equation}
In particular, the subscripted expressions indicates that the allowed representations are
restricted to $l_1>i$ and $l_2>j$.  If the subscript is absent, then the corresponding Dynkin
label need only be non-negative.

Finally, this allows us to write the $\mathbb Z_3$ singlet content of the branched $\mathrm{SU}(4)$
representations in (\ref{eq:branching}) as
\begin{eqnarray}
(0,n,0)&\longrightarrow&[0,0]_0(n),\nn\\
(1,n,0)&\longrightarrow&[2_{>0},-1]_{-1}(n)\oplus[0,0]_{-1}(n),\nn\\
(2,n,0)&\longrightarrow&[4_{>1},-2]_{-2}(n)\oplus[2_{>0},-1]_{-2}(n)\oplus[0,0]_{-2}(n),\nn\\
(1,n,1)&\longrightarrow&[2_{>0},-1]_0(n)\oplus[1_{>0},1_{>0}]_0(n)\oplus[0,0]_0(n)
\oplus[-1,2_{>0}]_0(n),\nn\\
(2,n,1)&\longrightarrow&[4_{>1},-2]_{-1}(n)\oplus[3_{>1},0_{>0}]_{-1}(n)\oplus[2_{>0},-1]_{-1}(n)
\oplus[1_{>0},1_{>0}]_{-1}(n)\nn\\
&&\oplus[0,0]_{-1}(n)\oplus[-1,2_{>0}]_{-1}(n),\nn\\
(2,n,2)&\longrightarrow&[4_{>1},-2]_0(n)\oplus[2_{>0},-1]_0(n)\oplus[3_{>1},0_{>0}]_0(n)
\oplus[2_{>1},2_{>1}]_0(n)\oplus[1_{>0},1_{>0}]_0(n)\nn\\
&&\oplus[0,0]_0(n)\oplus[0_{>0},3_{>1}]_0(n)
\oplus[-1,2_{>0}]_0(n)\oplus[-2,4_{>1}]_0(n).
\label{eq:Z3singlets}
\end{eqnarray}
%

\section{$\mathcal N=2$ multiplet structure}
\label{app:multiplets}

The grouping of the $S^5/\mathbb Z_3$ KK spectrum shown in Table~\ref{tbl:S5Z3spectrum}
into $\mathcal N=2$ representations proceeds by splitting off one set of $\mathrm{SU}(3)$
representations at a time, where the five possible sets are given in (\ref{eq:fivesets}).  Starting
with $[0,0]$, we find the fields
\begin{equation}
\varphi^{(1)},\quad\lambda^{(1)},\quad\varphi^{(2)},\quad A_\mu^{(1)},\quad\lambda^{(3)},
\quad\varphi^{(4)},
\end{equation}
which is suggestive of a vector multiplet.  A more careful consideration of the
quantum numbers shows that this in fact fills out Vector Multiplet I, in the language of
\cite{Ceresole:1999zs,Ceresole:1999ht,Eager:2012hx}.  A similar consideration of
the $[-2,-2]$ set gives the fields
\begin{equation}
\varphi^{(4)},\quad\varphi^{(5)},\quad\varphi^{(6)},\quad\lambda^{(5)},\quad\lambda^{(6)},
\quad A_\mu^{(3)},
\end{equation}
which fills out Vector Multiplet II.  The remaining sets of representations are slightly
more challenging to disentangle, as they give rise to a combination of multiplets.  The
$[1,-2]\sim[-2,1]$ set corresponds to Gravitino Multiplets I and III, the $[-4,2]\sim[-1,-1]
\sim[2,-4]$ set spits into a Graviton Multiplet and Vector Multiplets III and IV, and the
$[-3,0]\sim[0,-3]$ set corresponds to Gravitino Multiplets II and IV.  The results are presented
in Tables~\ref{tbl:GrM}--\ref{tbl:VM3}, in a similar format as those in
\cite{Ceresole:1999zs,Ceresole:1999ht}.

\begin{table}[tp]
\centering
\begin{tabular}{|c|c|c|c|c|c|r|l|}
 \hline
 & & &$(s_{1},s_{2})$  & $E_0$  & $R$-symm. & Field & $\mathrm{SU}(3)$ symbol \\
\hline
$\diamond$ & $\star$&$\bar\star$  & $(1,1)$  & $p+2$ & $r$&$h_{\mu\nu}$ & $[-1,-1]_0$\\
$\diamond$ & $\star$&$\bar\star$ & $(1,1/2)$  & $p+3/2$ & $r-1$&$\psi^{(1)}_{\mu}$ & $[-1,-1]_{-1}$\\
$\diamond$ & $\star$&$\bar\star$ & $(1/2,1)$  & $p+3/2$ & $r+1$&$\psi^{(1)}_{\mu}$ & $[-1,-1]_1$\\
& &$\bar\star$& $(1/2,1)$  & $p+5/2$ & $r-1$&$\psi^{(2)}_{\mu}$ & $[-1,-1_{>0}]_{-1}$\\
&$\star$&& $(1,1/2)$  & $p+5/2$ & $r+1$&$\psi^{(2)}_{\mu}$ & $[-1_{>0},-1]_1$\\
\cline{4-6}
$\diamond$ & $\star$&$\bar\star$ & ${(1/2,1/2)}$  & ${p+1}$ & $r$&${{A^{(1)}_{\mu}}}$ & $[-1,-1]_0$\\
\cline{4-6}
& $\star$& &$(1/2,1/2)$  & $p+2$ & $r+2$&$A^{(2)}_{\mu}$ & $[-1_{>0},-1]_2$\\
& &$\bar\star$& $(1/2,1/2)$  & $p+2$ & $r-2$&$A^{(2)}_{\mu}$ & $[-1,-1_{>0}]_{-2}$\\
& & &$(1/2,1/2)$  & $p+3$ & $r$&$A^{(3)}_{\mu}$ & $[-1_{>0},-1_{>0}]_0$\\
&$\star$& &$(1,0)$  & $p+2$ & $r$&$A^{(2)}_{\mu\nu}$ & $[-1_{>0},-1]_0$\\
& &$\bar\star$ &$(0,1)$  & $p+2$ & $r$&$A^{(2)}_{\mu\nu}$ & $[-1,-1_{>0}]_0$\\
&$\star$ &&$(1/2,0)$  & $p+3/2$ & $r+1$&$\lambda^{(3)}$ & $[-1_{>0},-1]_1$\\
&& $\bar\star$ & $(0,1/2)$  & $p+3/2$ & $r-1$&$\lambda^{(3)}$ & $[-1,-1_{>0}]_{-1}$\\
& & &$(1/2,0)$  & $p+5/2$ & $r-1$&$\lambda^{(5)}$ & $[-1_{>0},-1_{>0}]_{-1}$\\
& & &$(0,1/2)$  & $p+5/2$ & $r+1$&$\lambda^{(5)}$ & $[-1_{>0},-1_{>0}]_1$\\
& & &$(0,0)$  & $p+2$ & $r$&$\varphi^{(4)}$ & $[-1_{>0},-1_{>0}]_0$\\
\hline
\end{tabular}
\caption{\label{tbl:GrM} Graviton Multiplet, $\mathcal D(p+1,\ft12,\ft12;[-1,-1]_0)$, formed
by $[-1,-1]$ representations.}
\end{table}

\begin{table}[tp]
\centering
\begin{tabular}{|c|c|c|c|c|c|r|l|}
 \hline
& & &$(s_{1},s_{2})$  & $E_0$  & $R$-symm. & Field & $\mathrm{SU}(3)$ symbol\\
\hline
& $\star$ &$\bar\star$ & $(1,1/2)$  & $p+3/2$ & $r$&$\psi^{(1)}_{\mu}$ & $[1_{>0},-2]_{-1}$\\
& $\star$ &$\bar\star$& $(1/2,1/2)$  & $p+1$ & $r+1$&$A^{(1)}_{\mu}$ & $[1_{>0},-2]_{0}$\\
& &$\bar\star$ & $(1/2,1/2)$  & $p+2$ & $r-1$&$A^{(2)}_{\mu}$ & $[1_{>0},-2_{>0}]_{-2}$\\
$\bullet$ & $\star$ &$\bar\star$& $(1,0)$  & $p+1$ & $r-1$&$A^{(1)}_{\mu\nu}$ & $[1,-2]_{-2}$\\
&$\star$& &$(1,0)$ & $p+2$ & $r+1$&$A^{(2)}_{\mu\nu}$ & $[1_{>1},-2]_{0}$\\
\cline{4-6}
$\bullet$ & $\star$ &$\bar\star$& $(1/2,0)$  & $p+1/2$ & $r$&$\lambda^{(1)}$ & $[1,-2]_{-1}$\\
\cline{4-6}
$\bullet$ & &$\bar\star$ & $(1/2,0)$  & $p+3/2$ & $r-2$&$\lambda^{(2)}$ & $[1,-2_{>0}]_{-3}$\\
& &$\bar\star$& $(0,1/2)$  & $p+3/2$ & $r$&$\lambda^{(3)}$ & $[1_{>0},-2_{>0}]_{-1}$\\
&$\star$&& $(1/2,0)$  & $p+3/2$ & $r+2$&$\lambda^{(3)}$ & $[1_{>1},-2]_{1}$\\
& & &$(1/2,0)$  & $p+5/2$ & $r$&$\lambda^{(5)}$ & $[1_{>1},-2_{>0}]_{-1}$\\
$\bullet$ && $\bar\star$ & $(0,0)$  & $p+1$ & $r-1$&$\varphi^{(2)}$ & $[1,-2_{>0}]_{-2}$\\
& & &$(0,0)$  & $p+2$ & $r+1$&$\varphi^{(4)}$ & $[1_{>1},-2_{>0}]_{0}$\\
\hline
\end{tabular}
\caption{\label{tbl:Gino1} Gravitino Multiplet I, $\mathcal D(p+\ft12,\ft12,0;[1,-2]_{-1})$,
formed by $[1,-2]$ representations.}
\end{table}

\begin{table}[tp]
\centering
\begin{tabular}{|c|c|c|c|r|l|}
 \hline
& $(s_{1},s_{2})$  & $E_0$  & $R$-symm. & Field & $\mathrm{SU}(3)$ symbol\\
\hline
$\star$& $(1,1/2)$  & $p+5/2$ & $r$&$\psi^{(2)}_{\mu}$ & $[-3,0]_1$\\
$\star$& $(1/2,1/2)$  & $p+2$ & $r+1$&$A^{(2)}_{\mu}$ & $[-3,0]_2$\\
&$(1/2,1/2)$  & $p+3$ & $r-1$&$A^{(3)}_{\mu}$ & $[-3,0_{>0}]_0$\\
$\star$& $(1,0)$  & $p+2$ & $r-1$&$A^{(2)}_{\mu\nu}$ & $[-3,0]_0$\\
$\star$& $(1,0)$  & $p+3$ & $r+1$&$A^{(3)}_{\mu\nu}$ & $[-3,0]_2$\\
\cline{2-4}
$\star$& $(1/2,0)$  & $p+3/2$ & $r$&$\lambda^{(3)}$ & $[-3,0]_1$\\
\cline{2-4}
& $(1/2,0)$  & $p+5/2$ & $r-2$&$\lambda^{(5)}$ & $[-3,0_{>0}]_{-1}$\\
& $(0,1/2)$  & $p+5/2$ & $r$&$\lambda^{(5)}$ & $[-3,0_{>0}]_1$\\
$\star$& $(1/2,0)$  & $p+5/2$ & $r+2$&$\lambda^{(4)}$ & $[-3,0]_3$\\
& $(1/2,0)$  & $p+7/2$ & $r$&$\lambda^{(6)}$ & $[-3,0_{>0}]_1$\\
& $(0,0)$  & $p+2$ & $r-1$&$\varphi^{(4)}$ & $[-3,0_{>0}]_0$\\
& $(0,0)$  & $p+3$ & $r+1$&$\varphi^{(5)}$ & $[-3,0_{>0}]_2$\\
\hline
\end{tabular}
\caption{\label{tbl:Gino2} Gravitino Multiplet II, $\mathcal D(p+\ft32,\ft12,0;[-3,0]_1)$,
formed by $[-3,0]$ representations.}
\end{table}

\begin{table}[tp]
\centering
\begin{tabular}{|c|c|c|c|c|c|c|c|r|l|}
\hline
& & &&& $(s_{1},s_{2})$  & $E_0$  & $R$-symm. & Field & $\mathrm{SU}(3)$ symbol\\
\hline
$\diamond$&&&$\star$&$\bar\star$&$(1/2,1/2)$&$p+1$&$r$&$A^{(1)}_{\mu}$&$[0_{>0},0_{>0}]_0$\\
$\diamond$&$\bullet$&&$\star$&$\bar\star$&$(1/2,0)$&$p+1/2$&$r-1$&$\lambda^{(1)}$
    &$[0,0_{>0}]_{-1}$\\
$\diamond$&&$\bar\bullet$&$\star$&$\bar\star$&$(0,1/2)$&$p+1/2$&$r+1$&$\lambda^{(1)}$
    &$[0_{>0},0]_1$\\
& & &&$\bar\star$ & $(0,1/2)$  & $p+3/2$ & $r-1$&$\lambda^{(3)}$ & $[0_{>0},0_{>1}]_{-1}$\\
& &  &$\star$& & $(1/2,0)$  & $p+3/2$ & $r+1$&$\lambda^{(3)}$ & $[0_{>1},0_{>0}]_{1}$\\
\cline{6-8}
$\diamond$&$\bullet$&$\bar\bullet$ &$\star$&$\bar\star$&$(0,0)$&$p$&$r$&$\varphi^{(1)}$
    &$[0,0]_{0}$\\
\cline{6-8}
&$\bullet$ & &&$\bar\star$ & $(0,0)$  & $p+1$ & $r-2$&$\varphi^{(2)}$ & $[0,0_{>1}]_{-2}$\\
& &$\bar\bullet$ &$\star$& & $(0,0)$  & $p+1$ & $r+2$&$\varphi^{(2)}$ & $[0_{>1},0]_{2}$\\
& & &  && $(0,0)$  & $p+2$ & $r$&$\varphi^{(4)}$ & $[0_{>1},0_{>1}]_{0}$\\
\hline
\end{tabular}
\caption{\label{tbl:VM1} Vector Multiplet I, $\mathcal D(p,0,0;[0,0]_0)$,
formed by $[0,0]$ representations.}
\end{table}

\begin{table}[tp]
\centering
\begin{tabular}{|c|c|c|r|l|}
\hline
$(s_{1},s_{2})$  & $E_0$  & $R$-symm. & Field & $\mathrm{SU}(3)$ symbol\\
\hline
$(1/2,1/2)$  & $p+3$ & $r$\ \ &$A^{(3)}_{\mu}$ & $[-2,-2]_0$\\
$(1/2,0)$  & $p+5/2$ & $r-1$\ \ &$\lambda^{(5)}$ & $[-2,-2]_{-1}$\\
$(0,1/2)$  & $p+5/2$ & $r+1$\ \ &$\lambda^{(5)}$ & $[-2,-2]_1$\\
$(0,1/2)$  & $p+7/2$ & $r-1$\ \ &$\lambda^{(6)}$ & $[-2,-2]_{-1}$\\
$(1/2,0)$  & $p+7/2$ & $r+1$\ \ &$\lambda^{(6)}$ & $[-2,-2]_1$\\
\cline{1-3}
$(0,0)$  & $p+2$ & $r$\ \ &$\varphi^{(4)}$ & $[-2,-2]_0$\\
\cline{1-3}
$(0,0)$  & $p+3$ & $r-2$\ \ &$\varphi^{(5)}$ & $[-2,-2]_{-2}$\\
$(0,0)$  & $p+3$ & $r+2$\ \ &$\varphi^{(5)}$ & $[-2,-2]_2$\\
$(0,0)$  & $p+4$ & $r$\ \ &$\varphi^{(6)}$ & $[-2,-2]_0$\\
\hline
\end{tabular}
\caption{\label{tbl:VM2} Vector Multiplet II, $\mathcal D(p+2,0,0;[-2,-2]_0)$,
formed by $[-2,-2]$ representations.}
\end{table}

\begin{table}[tp]
\centering
\begin{tabular}{|c|c|c|c|c|r|l|}
\hline
& & $(s_{1},s_{2})$  & $E_0$  & $R$-symm. & Field & $\mathrm{SU}(3)$ symbol\\
\hline
 & $\star$ & $(1/2,1/2)$  & $p+2$ & $r$&$A^{(2)}_{\mu}$ & $[-1,-1_{>0}]_{-2}$\\
$\bar\bullet$ &$\star$  & $(0,1/2)$  & $p+3/2$ & $r+1$&$\lambda^{(2)}$ & $[-1,-1]_{-1}$\\
 & $\star$ & $(1/2,0)$  & $p+3/2$ & $r-1$&$\lambda^{(3)}$ & $[-1,-1_{>0}]_{-3}$\\
 & $\star$ & $(1/2,0)$  & $p+5/2$ & $r+1$&$\lambda^{(4)}$ & $[-1,-1_{>0}]_{-1}$\\
 &   & $(0,1/2)$  & $p+5/2$ & $r-1$&$\lambda^{(5)}$ & $[-1,-1_{>1}]_{-3}$\\
 \cline{3-5}
$\bar\bullet$ & $\star$ & $(0,0)$  & $p+1$ & $r$&$\varphi^{(2)}$ & $[-1,-1]_{-2}$\\
\cline{3-5}
$\bar\bullet$ & $\star$ & $(0,0)$  & $p+2$ & $r+2$&$\varphi^{(3)}$ & $[-1,-1]_0$\\
 & & $(0,0)$  & $p+2$ & $r-2$&$\varphi^{(4)}$ & $[-1,-1_{>1}]_{-4}$\\
 & & $(0,0)$  & $p+3$ & $r$&$\varphi^{(5)}$ & $[-1,-1_{>1}]_{-2}$\\
\hline
\end{tabular}
\caption{\label{tbl:VM3} Vector Multiplet III, $\mathcal D(p+1,0,0;[-1,-1]_{-2})$,
formed by $[-1,-1]$ representations.}
\end{table}

Multiplet shortening in the tables are indicated in the first few columns.
Massless (conserved) multiplets are marked by diamonds, chiral multiplets by
bullets and semi-long multiplets by stars.  Bars on top of the symbols indicates
anti-chiral or semi-long II shortening.  We have not included the corresponding tables
for Gravitino Multiplets III and IV and Vector Multiplet IV, as they are conjugate to
Gravitino Multiplets I and II, and Vector Multiplet III.  Note that Vector Multiplet II
is never shortened.

The $\mathrm{SU}(3)$ symbols are defined in Appendix~\ref{app:Z3trunc}, and represent
a set of triality zero $\mathrm{SU}(3)$ representations that show up in the expansion at
KK level $p$.  Since all states within a given multiplet transform identically under
$\mathrm{SU}(3)$, they share a common set of representations generated by $[a,b]$.
Note, however, that the $R$-charges are shifted as appropriate for different states in
the multiplet.

From these tables, we can see the connection between the subscripted
restrictions on the Dynkin labels (`$>0$' or `$>1$') and multiplet
shortening.  For example, consider the Graviton Multiplet of
Table~\ref{tbl:GrM}.  As indicated in Table~\ref{tbl:KKshort}, this
shortens to a massless graviton multiplet at level $p=2$.  Since
this transforms as a singlet of $\mathrm{SU}(3)$, or equivalently as
the $(0,0)$ representation, all `$>0$' subscripted states are absent
in this case, leaving only the states marked by diamonds.  At levels
$p=3l+2$, the Graviton Multiplet shortens into semi-long
representations at the extremes of the $\mathrm{SU}(3)$ sequence,
given by $(3l,0)$ and $(0,3l)$.  In the former semi-long I case, the
states in Table~\ref{tbl:GrM} with subscript `$>0$' in the second
position are absent, leaving only the states marked by stars.
Further examination of the remaining multiplets demonstrates that
the shortening conditions are all consistent with the restrictions
on the KK states, as guaranteed by supersymmetry.


\end{document}